\documentclass[twocolumn,
superscriptaddress,
preprintnumbers,
amsmath,
amssymb,
]{revtex4-1}

  \usepackage{grffile}
  \usepackage{graphicx}
  \usepackage{color}
  \usepackage{dcolumn}
  \usepackage{bm}
\begin{document}

\title{Interface induced spin-orbit interaction in silicon quantum dots and prospects for scalability}

\author{Rifat Ferdous}
\affiliation{Network for Computational Nanotechnology, Purdue University, West Lafayette, IN 47907, USA}

\author{Kok W. Chan}
\affiliation{Centre for Quantum Computation and Communication Technology, School of Electrical Engineering and Telecommunications, The University of New South Wales, Sydney, New South Wales 2052, Australia}

\author{Menno Veldhorst}
\affiliation{QuTech and Kavli Institute of Nanoscience, TU Delft, Lorentzweg 1, 2628CJ Delft, the Netherlands}

\author{J.C.C. Hwang}
\affiliation{Centre for Quantum Computation and Communication Technology, School of Electrical Engineering and Telecommunications, The University of New South Wales, Sydney, New South Wales 2052, Australia}

\author{C. H. Yang}
\affiliation{Centre for Quantum Computation and Communication Technology, School of Electrical Engineering and Telecommunications, The University of New South Wales, Sydney, New South Wales 2052, Australia}

\author{Gerhard Klimeck}
\affiliation{Network for Computational Nanotechnology, Purdue University, West Lafayette, IN 47907, USA}

\author{Andrea Morello}
\affiliation{Centre for Quantum Computation and Communication Technology, School of Electrical Engineering and Telecommunications, The University of New South Wales, Sydney, New South Wales 2052, Australia}

\author{Andrew S. Dzurak}
\affiliation{Centre for Quantum Computation and Communication Technology, School of Electrical Engineering and Telecommunications, The University of New South Wales, Sydney, New South Wales 2052, Australia}

\author{Rajib Rahman}
\affiliation{Network for Computational Nanotechnology, Purdue University, West Lafayette, IN 47907, USA}

\date{\today}

\begin{abstract}
We identify the presence of monoatomic steps at the Si/SiGe or Si/SiO$_2$ interface as a dominant source of variations in the dephasing time of Si quantum dot (QD) spin qubits. First, using atomistc tight-binding calculations we show that the g-factors and their Stark shifts undergo variations due to these steps. We compare our theoretical predictions with experiments on QDs at a Si/SiO$_2$ interface, in which we observe  significant differences in Stark shifts between QDs in two different samples. We also experimentally observe variations in the $g$-factors of one-electron and three-electron spin qubits realized in three neighboring QDs on the same sample, at a level consistent with our calculations. The dephasing times of these qubits also vary, most likely due to their varying sensitivity to charge noise, resulting from different interface conditions. More importantly, from our calculations we show that by employing the anisotropic nature of the spin-orbit interaction (SOI) in a Si QD, we can minimize and control these variations. Ultimately, we predict that the dephasing times of the Si QD spin qubits will be anisotropic and can be improved by at least an order of magnitude, by aligning the external DC magnetic field towards specific crystal directions.  

\end{abstract}

\maketitle


A scalable quantum computing architecture requires reproducibility and control over key qubit properties, such as resonance frequency, coherence time, etc. Variability in such parameters among qubits of a large-scale quantum computer would necessitate individual qubit characterization and control\cite{lieven_arxiv_2016}, while excessive variability could even make scaling impractical. In case of significant variability in the dephasing time, the performance of a large-scale quantum computer might be limited by the fidelity of the qubit that decoheres the fastest. 

Spin qubits hosted in silicon (Si) quantum dots (QD)\cite{loss_pra_1998} have been showing promise as a potential building block for a large-scale quantum computer\cite{zwaneburg_revmdphys_2013}, because of their compatibility with already existing CMOS technology and the long coherence times available due to the presence of negligible nuclear spins in isotopically purified $^{28}$Si\cite{itoh_mrscomm_2014}. Single\cite{HRL_2011,kim_nature_2014,wu_pnas_2014,veldhrost_natnano_2014,kawakami_natnano_2014,HRL_exchange_only} and two qubit\cite{veldhorst_nature_2015} gates have been demonstrated already. To move forward with increasing numbers of qubits\cite{zajac_pra_2016,cjones_arxiv_2016,veldhorst_arxiv_2016,lieven_arxiv_2016}, we have to study possible sources that can  cause variations in the coherence time and limit the performance of these qubits. 


In this letter, we provide a microscopic understanding of the dephasing time $T_2^*$ of Si QD spin qubits. We show that electrical noise modulates the electron $g$-factor through spin-orbit interaction (SOI) and causes dephasing. Moreover, the atomic scale details of the interface controls the sensitivity of the g-factor to the electric field or noise and hence introduce variability in the $T_2^*$ times. We experimentally observe variations in the $g$-factors, their gate voltage dependence and $T_2^*$ times among spin qubits hosted in gate-defined quantum dots formed at a Si/SiO$_2$ interface. Finally we predict that, due to the anisotropic nature of the SOI in Si QDs, the $T_2^*$ times will be anisotropic and hence can be improved and their variability can be reduced as well by choosing the appropriate direction of the external magnetic field.

The energy levels of interest in a Si QD for qubit operations are two low lying conduction band valley states $v_-$ and $v_+$, each split in two spin levels in the presence of a DC magnetic field, $B_{\textup{ext}}$. Until recently, the main focus has been to increase the valley splitting energy to create well isolated states. In metal-oxide-semiconductor based Si QDs, this is now routinely achieved\cite{veldhrost_natnano_2014,yang_natcomm_2013}. However, it turns out that the spin splitting ($E_{\textup{ZS}}^{\pm}=g_{\pm}\mu_\textup{B} B_{\textup{ext}}$, where $\mu_\textup{B}$ is the Bohr magneton) and also the dephasing time $T_2^*$  are valley dependent\cite{kawakami_natnano_2014,veldhorst_prbrap_2015,Scarlino_prl_2015,Scarlino_prb_2017,rifat_arxiv_2016} and, as we will show experimentally, is sample-to-sample dependent. 


In a Si quantum well or dot, the presence of structure inversion asymmetry (SIA) introduces the Rashba SOI\cite{golub_prb_2004,nestoklon_prb_2006,nestoklon_prb_2008}. Though it is known that due to the lack of bulk inversion asymmetry (BIA), the Dresselhaus SOI is absent from bulk Si, interface inversion asymmetry (IIA) contributes a Dresselhaus-like term in interface confined structures in Si\cite{golub_prb_2004,nestoklon_prb_2006,nestoklon_prb_2008}. Both the Rashba and Dresselhaus SOI modify the electron $g$-factors in a Si QD, and enable the Stark shift of the $g$-factors through gate voltage tuning\cite{veldhrost_natnano_2014,yang_natcomm_2013,veldhorst_prbrap_2015}. The different sign of the Rashba ($\alpha_{\pm}$) and Dresselhaus coefficients ($\beta_{\pm}$) results in different g-factors among the two valley states\cite{veldhorst_prbrap_2015}. The Dresselhaus contribution is usually much stronger than the Rashba SOI\cite{nestoklon_prb_2008,rifat_arxiv_2016}, and dominates the $g$-factor renormalization \cite{rifat_arxiv_2016}. These SOI effects also make the qubits susceptible to electrical noise.


In a Si QD with a strong vertical electric field, the electrons are usually confined to only one interface. A monoatomic shift in the location of this interface results in a sign inversion of the Dresselhaus coefficient ($\beta$), while the Rashba coefficient ($\alpha$) remains unchanged\cite{golub_prb_2004,nestoklon_prb_2006,nestoklon_prb_2008}. In practice, Si/SiGe or Si/SiO$_2$ interfaces certainly contain roughness, i.e monoatomic steps\cite{steps_justify,Friesen_steps,kharche_apl_2007}. A non-ideal interface with monoatomic steps can be thought of as multiple smooth interface regions, where interfaces of neighboring regions are shifted by one atomic layer with respect to each other. Thus the neighboring regions will have opposite signs of $\beta$. An electron wavefunction spread over multiple such regions will witness multiple local $\beta$s and the effective $\beta$ will be a weighted average. Thus the presence of interface steps can change both the sign and magnitude of the effective Dresselhaus contribution to the electron $g$-factors in a Si QD\cite{rifat_arxiv_2016}. To accurately understand these atomic-scale physics of the interface, here we use spin resolved $sp^3d^5s^*$ tight-binding model, where the effects of the SOI comes out automatically based on the atomic arrangement of the QD, without any pre-assumption about the Rashba or Dresselhaus SOI.



\begin{figure}[htbp]
\includegraphics[width=\linewidth]{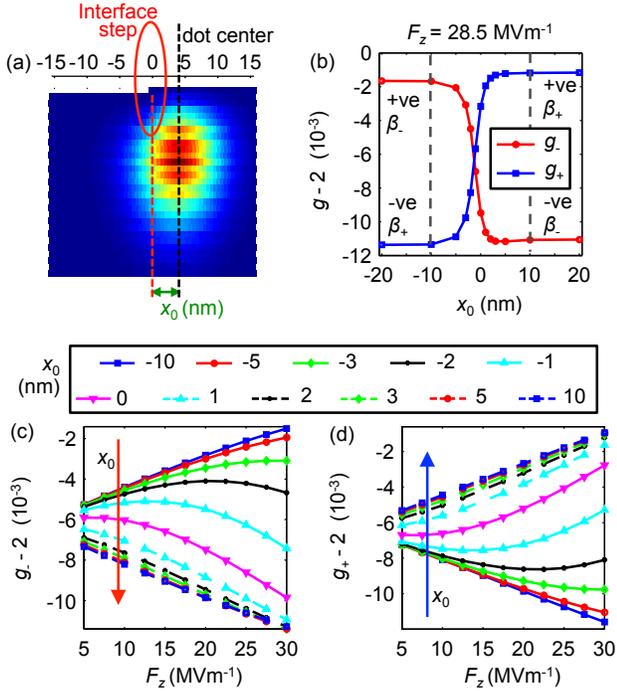}
\caption{{Effect of interface steps on g-factors and their Stark shifts in a Si QD from atomistic tight-binding calculation.} (a)  An electron wavefunction subject to an interface step. (b) Variation in the g-factors for both valley states ($g_-$ and $g_+$), as a function of $x_0$ for vertical electric field $F_z=$28.5 MVm$^{-1}$\cite{yang_natcomm_2013}. $F_z$ dependence of (c) $g_-$  and (d) $g_+$ for various $x_0$. }
\vspace{0cm}
\label{fi1}
\end{figure}

Fig.\ 1 shows how a monoatomic step at the interface of a Si QD can affect the $g$-factors of the valley states and their electric field dependence, with an external magnetic field along the [110] crystal orientation, from atomistic tight-binding simulations. An electron wavefunction near an interface step is shown in Fig.\ 1a. The distance between the dot center and the location of the edge of the interface step is denoted by $x_0$. The dot radius is around 10 nm. So for $x_0<$-10 nm the dot is completely on the left side of the step and has different $g$-factors ($g_->g_+$) compared to that ($g_+>g_-$) for $x_0>$10 nm, when the dot is completely on the right side of the step, as seen in Fig.\ 1b. For -10 nm$<x_0<$10 nm, the $g$-factors are a weighted average of those of the two sides based on the dot location. To understand this atomistic calculation we use an analytic effective mass model that relates $g_{\pm}$ in a Si QD, with the Rashba and Dresselhaus SOI\cite{veldhorst_prbrap_2015,rifat_arxiv_2016}. We briefly summarize this model in the Supplemental Material\cite{supplementary}. For $B_{\textup{ext}}$ along the [110] crystal orientation
\begin{align}
\delta g_{\pm}^{[110]}=2\frac{\left| e \right|\left\langle z \right\rangle }{\mu_\textup{B} \hbar }\left(-{\alpha }_{\pm }+{{\beta }_{\pm }}\right)  
\label{equ4}
\end{align}  
Here, $\left| e \right|$ is the electron charge, $\left\langle z \right\rangle$ is the spread of the electron wavefunction along the vertical direction z ([001]) and $\hbar$ is the reduced Planck constant. Now, in a Si QD, $\beta>>\alpha$\cite{nestoklon_prb_2008,rifat_arxiv_2016}, and so
\begin{align}
\delta g_{\pm}^{[110]}\approx2\frac{\left| e \right|\left\langle z \right\rangle }{\mu_\textup{B} \hbar }{{\beta }_{\pm }}  
\label{equ5}
\end{align}


 As previously discussed, $\beta$ has a different sign between the two sides of the step. When the location of the dot changes with respect to the step, the weighted average of the positive and negative $\beta$s change, which changes the g-factors.

Figs. 1c and 1d show that the Stark shift of the $g$-factors, as a function of the confining vertical electric field $F_z$, for both valley states are also affected by the presence of an interface step. The differential change in the $g$-factors with electric field, $\frac{dg_{\pm}}{dF_z}$, can vary in both sign and magnitude depending on the location of the step with respect to the dot center. This behavior can also be explained by equation \ref{equ5}, with the change in $\beta$ near an interface step. For example in Fig.\ 1c, for $x_0\approx$ -10 nm, the dot is completely on the left side of the step, where the $v_-$ valley state has positive $\beta$. Thus an increase in $\beta_-$ with increasing $F_z$ increases $g_-$ as well, hence a positive $\frac{dg_-}{dF_z}$. On the other hand, when the dot is completely on the right side of the step, at $x_0\approx$ 10 nm, $\beta_-$ is negative. Thus increasing $F_z$ increases $|\beta_-|$ but decreases $g_-$ and thus results in a negative $\frac{dg_-}{dF_z}$. For -10 nm$<x_0<$10 nm, $\frac{dg_-}{dF_z}$ changes gradually with $x_0$. We see a similar but opposite change for $g_+$ in Fig.\ 1d.   


\begin{figure}[htbp]
\includegraphics[trim=0 0 0 0, clip, width=\linewidth]{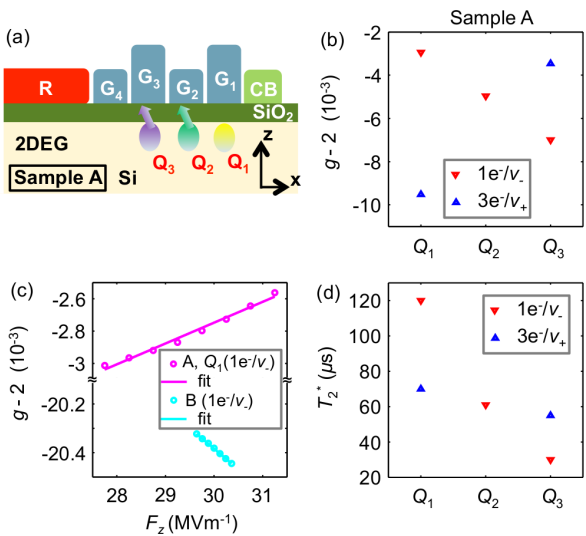}
\caption{{Schematic diagram of the experimental sample and observed dot-to-dot variations.} (a) Cross sectional schematic of sample A. CB acts as a lateral confinement gate in the formation of quantum dots under gates G$_1$, G$_2$, and G$_3$. G$_4$ is used as a tunnel barrier for loading/unloading of electrons from the 2DEG formed under the reservoir (R) gate. (b) Variation in the $g$-factors, both one-electron ($g_-$) and three-electron ($g_+$), among three neighboring quantum dots ($Q_1$, $Q_2$, $Q_3$) formed at the Si/SiO$_2$ interface in sample A. (c) One-electron Stark shift of $Q_1$ from sample A and one QD from sample B plotted together as a function of the vertical electric field, $F_z$. Note that both samples were measured in different dilution fridges and there is an unknown $B_{\textup{ext}}$ offset in sample B, contributing to larger discrepancy in its $g$-factor from 2. (d) Observed variations in the dephasing times among qubits in sample A.}
\vspace{0cm}
\label{fi2}
\end{figure}


Similar variations in the $g$-factors, and their gate voltage dependence, are measured in gate-defined quantum dots formed at a Si/SiO$_2$ interface for two different samples (A and B) with similar architecture. In sample A we operated spin qubits in three neighboring QDs while in sample B we studied a qubit in a single QD. In Fig.\ 2a, we show a schematic diagram of one of the devices (sample A) measured experimentally. In sample A, we observe variations in one-electron and three-electron $g$-factors among three neighboring QDs ($Q_1$, $Q_2$, $Q_3$), as shown in Fig.\ 2b. We understand that the one-electron (three-electron) qubit corresponds to an electron occupying the lower (higher) energy valley state $v_-$ ($v_+)$\cite{veldhorst_prbrap_2015}. We could not achieve three-electron spin resonance for $Q_2$ as it was strongly coupled to the other dots. Fig.\ 2c shows the Stark shift of $Q_1$ on sample A and one QD on another sample B. We see here that the $g_-$ of $Q_1$ has opposite dependence on $F_z$ compared to that of the QD in sample B. These observed variations in both the Stark shifts and the $g$-factors qualitatively agree with the theoretically predicted variations shown in Fig.\ 1. We therefore conclude that these experimentally observed variations are primarily due to different interface conditions associated with each of the QDs.



We also observe variations in the measured $T_2^*$ times for both valley states of the three QDs, as shown in Fig.\ 2d. The $T_2^*$ times were extracted by performing Ramsey experiments on all of the qubits in sample A, as shown in the Supplemental Material\cite{supplementary}. 

The observed variations in $T_2^*$ can be explained from the measured (Fig.\ 2c) and predicted (Fig.\ 1c,d) variations in $|\frac{dg_{\pm}}{dF_z}|$ with interface steps. The larger the $|\frac{dg_{\pm}}{dF_z}|$, the more sensitive the qubit is to electrical noise. The dephasing time due to nuclear spin fluctuations is given in refs.\ \cite{RHanson_revmodp_2007,merkulov_prb_2002} and in our samples, which employ an isotopically enriched $^{28}$Si substrate, these times are very long. In the absence of nuclear spin, we can relate $T_2^*$ times with electrical noise in a similar way,

\begin{align}
T_2^*=\frac{\sqrt{2}\hbar}{\Delta F_z |\frac{dg}{dF_z}|\mu_\textup{B} B_{\textup{ext}} }
\label{equ1}
\end{align} 
Here, $\Delta F_z$ is the standard deviation of the electric field fluctuation seen by the dot, due to electrical noise on the gate. For $Q_1$ in sample A, we extract $|\frac{dg_-}{dF_z}|\approx1.1\times10^{-10}$ mV$^{-1}$ from Fig.\ 2c and estimate $\Delta F_z\approx870$ Vm$^{-1}$ for $T_2^*=120$ $\mu s$. When we compare the $T_2^*$ times between the two valley states of $Q_1$, we see $T_2^*(v_-,Q_1)\approx 1.7$ $T_2^*(v_+,Q_1)$ and from ref.\ \cite{veldhorst_prbrap_2015} we find, $|\frac{dg_+^{Q_1}}{dF_z}|\approx2.2$ $|\frac{dg_-^{Q_1}}{dF_z}|$. This comparison also shows how $\frac{1}{T_2^*}$ almost linearly depends on $|\frac{dg_{\pm}}{dF_z}|$ as predicted by equation \ref{equ1}. Thus variations in $|\frac{dg_{\pm}}{dF_z}|$ due to different interface conditions among different qubits will result in a variability in the dephasing time. 

The calculations of Fig.\ 1 and the experimental observations of Fig.\ 2 highlight the device-to-device variability issues that would require individual knowledge of each qubit, and impose a challenge to the implementation of a large scale quantum computer. Any possible way of reducing the variability is crucial to the scale up of Si QD spin qubits. Also an increase in $T_2^*$, regardless of the interface condition, will aid scalability. Next we investigate ways to improve these issues.

One obvious way to suppress these variabilities and gain more control over the g-factors, their tunability and the dephasing times, is to minimize interface roughness, which is a well known fabrication challenge. Here we propose an alternate approach. As predicted in ref.\ \cite{rifat_arxiv_2016}, the $g$-factors in a Si QD are anisotropic. We can study the anisotropy from a simplified expression\cite{rifat_arxiv_2016,supplementary},

\begin{align}
\delta g_{\pm}\approx2\frac{\left| e \right|\left\langle z \right\rangle }{\mu_\textup{B} \hbar }\left(-{{\alpha }_{\pm }}+{{\beta }_{\pm }}\sin 2\phi \right) 
\label{equ2}
\end{align}     

Here, $\phi$ is the angle of the external magnetic field with the [100] crystal orientation. From equation \ref{equ2} we see that the contribution of the Dresselhaus SOI is anisotropic. Thus by changing the direction of $B_{\textup{ext}}$ we can tune the Dresselhaus contribution. For $\phi=0^{\circ}/90^{\circ}/180^{\circ}/270^{\circ}$, the Dresselhaus contribution will be negligible. For example, when $B_{\textup{ext}}$ is along [100], $\phi=0^{\circ}$ and

\begin{align}
\delta g_{\pm}^{[100]}\approx-2\frac{\left| e \right|\left\langle z \right\rangle }{\mu_\textup{B} \hbar }{{\alpha }_{\pm }}  
\label{equ3}
\end{align} 
Comparing equation \ref{equ5} and \ref{equ3} we see that,
\begin{align}
\frac{\delta g_{\pm}^{[100]}}{\delta g_{\pm}^{[110]}}\approx\frac{{\alpha }_{\pm }}{{\beta }_{\pm }}  
\label{equ6}
\end{align}
As the effect of the monoatomic steps are more dramatic on $\beta_{\pm}$, the change in $g_{\pm}$ and $\frac{dg_{\pm}}{dF_z}$ with interface steps should be smaller for $B_{\textup{ext}}$ along [100] compared to that for [110]. Moreover, since $\beta_{\pm}>>\alpha_{\pm}$ \cite{rifat_arxiv_2016,nestoklon_prb_2008}, $\frac{dg_{\pm}}{dF_z}$ itself will be much smaller for [100]. 

\begin{figure}[htbp]
\includegraphics[width=\linewidth]{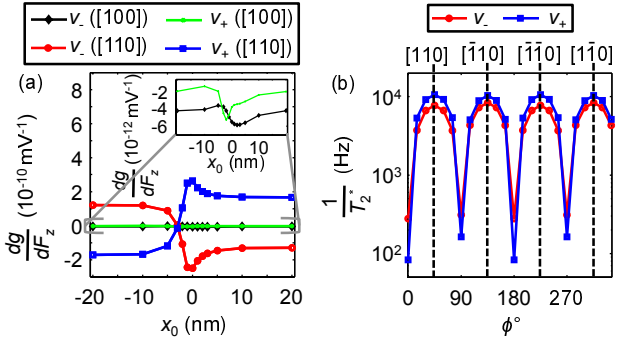}
\caption{(a) Change in $\frac{dg_{\pm}}{dF_z}$ as a function of $x_0$ with $B_{\textup{ext}}$ along [110] and [100] (inset), for $F_z$=28.5 MVm$^{-1}$, calculated using atomistic tight-binding model. (b) $\frac{1}{T_2^*}$ with respect to the direction of $B_{\textup{ext}}$, $\phi$ for $x_0$=6 nm, $F_z$=28.5 MVm$^{-1}$ and $B_{\textup{ext}}$=1.4015 T. }
\vspace{0cm}
\label{fi4}
\end{figure}

Fig.\ 3a shows variations in $\frac{dg_{\pm}}{dF_z}$ with $x_0$ for $B_{\textup{ext}}$ along [110] and [100]. Though there are variations in $\frac{dg_{\pm}}{dF_z}$ with $x_0$ for $B_{\textup{ext}}$ along [100], as shown by the inset of Fig.\ 3a, these variations and also $\frac{dg_{\pm}}{dF_z}$ themselves are negligible, when compared to that along [110]. Variation of $g_{\pm}$ with $x_0$, will also be negligible for $B_{\textup{ext}}$ along [100], as shown in the Supplemental Material\cite{supplementary}. Such phenomena will have a critical impact on the realization of a large scale quantum computer made of Si QDs. If the external magnetic field is along the [100] crystal orientation, all the qubits will have negligible variations in $g_{\pm}$, $\frac{dg_{\pm}}{dF_z}$, and consequently in $T_2^*$ even in the presence of varying interface conditions. Very small $|\frac{dg_{\pm}}{dF_z}|$ along [100] would also result in very long $T_2^*$ times. From Fig.\ 3a, we calculate the average $|\frac{dg_{\pm}}{dF_z}|$ over different values of $x_0$ and find that, $avg\left(|\frac{dg_{-}}{dF_z}\left([110]\right)|\right)\approx30$ $avg\left(|\frac{dg_{-}}{dF_z}\left([100]\right)|\right)$ and $avg\left(|\frac{dg_{+}}{dF_z}\left([110]\right)|\right)\approx50$ $avg\left(|\frac{dg_{+}}{dF_z}\left([100]\right)|\right)$. If a linear relationship between $\frac{1}{T_2^*}$ and $|\frac{dg_{\pm}}{dF_z}|$ holds, as predicted by equation \ref{equ1}, the $T_2^*$ times can be longer than 1 ms, provided that other noise sources do not contribute significantly.


In Fig.\ 3b, the angular dependence of $\frac{1}{T_2^*}$ for $x_0$=6 nm, is shown. From atomistic simulations we calculated $\frac{dg_{\pm}}{dF_z}$ for different $\phi$ and used equation \ref{equ1} to calculate $\frac{1}{T_2^*}$ for $\Delta F_z\approx870$ Vm$^{-1}$ and $B_{\textup{ext}}$=1.4 T. The anisotropy in Fig.\ 3b can be understood by connecting equation \ref{equ1} and \ref{equ2},

\begin{align}
\frac{1}{T_{2}^{*}\left( {{v}_{\pm }} \right)}=\Delta {{F}_{z}}\frac{2\left| e \right|B_{\textup{ext}}}{\sqrt{2}{{\hbar }^{2}}}\left| -\frac{d\left( \left\langle z \right\rangle {{\alpha }_{\pm }} \right)}{d{{F}_{z}}}+\sin 2\phi \frac{d\left( \left\langle z \right\rangle {{\beta }_{\pm }} \right)}{d{{F}_{z}}} \right| 
\label{equ7}
\end{align}

As $\frac{d\left( \left\langle z \right\rangle {{\beta }_{\pm }} \right)}{d{{F}_{z}}}\gg\frac{d\left( \left\langle z \right\rangle {{\alpha }_{\pm }} \right)}{d{{F}_{z}}}$ a large increase in $T_2^*$ is achievable by orientating $B_{\textup{ext}}$ along $[100]$/$[010]$/$[\bar{1}00]$/$[0\bar{1}0]$.


Now, a decrease in $|\frac{dg_{\pm}}{dF_z}|$ would also mean a reduced tunability of the $g$-factors, which is necessary for selective addressing of individual qubits. However, an increase in $T_2^*$ times will result in a narrower electron spin resonance (ESR) linewidth, $\delta f_{\textup{FWHM}}=\frac{2 \sqrt{\ln2}}{\pi T_2^*}$\cite{kawakami_natnano_2014}, which would then require a smaller difference in $g$-factors between qubits to individually address them. As the $g$-factors and $\frac{dg_{\pm}}{dF_z}$ have lesser variations with interface roughness along [100], the $g$-factor difference between qubits can be more controlled. Moreover, by tuning the direction of the external magnetic field a desired trade off between large $|\frac{dg_{\pm}}{dF_z}|$ and large $T_2^*$ times can be achieved. 

 Orienting the magnetic field along the [100] crystal orientation results in a Dresselhaus SOI with only off-diagonal components \cite{supplementary}. Therefore, electric field fluctuations, to first order, contribute to spin dephasing through the weaker Rashba SOI, ensuring a long $T_2^*$ time. At the same time, a resonant oscillating electric field can induce electric dipole spin resonance (EDSR) through the off-diagonal Dresselhaus coupling. Since $T_2^*$ is long under these conditions, coherent operations can be expected even for relatively weak EDSR driving strength, and without invoking the use of micromagnets\cite{kawakami_natnano_2014}.  

To conclude, we experimentally observe variations in the electron $g$-factors and their sensitivity to electric field, which also leads to variability in the dephasing times among Si/SiO$_2$ quantum dot spin qubits. We identify that the source of the variation is the presence of random monoatomic steps at the interface. To gain control over these key qubit parameters one has to minimize interface roughness, which is challenging to achieve. We show here that even in the presence of interface steps we can control and minimize these variations by taking advantage of the anisotropic SOI in a Si QD. Importantly, we can increase $T_2^*$ times if we align the external magnetic field along the [100] crystal orientation, rather than along [110]. Pointing the  external B-field along [100] will also help to reduce the SOI induced dephasing in Si QD devices with integrated micro-magnets, as SOI also contributes to the $g$-factors in these devices\cite{rifat_arxiv_2016}.




%

\section*{ACKNOWLEDGMENT}
This work was supported by the U.S. Army Research Office (W911NF-13-1-0024,W911NF-12-0607), the Australian Research Council (CE11E0001017), and the NSW Node of Australian National Fabrication Facility. Computational resources on
nanoHUB.org, funded by the NSF grant EEC-0228390, were used.

\end{document}


\title{Supplementary Materials for `Interface induced spin-orbit interaction in silicon quantum dots and prospects for scalability'}

\author{Rifat Ferdous}
\affiliation{Network for Computational Nanotechnology, Purdue University, West Lafayette, IN 47907, USA}

\author{Kok W. Chan}
\affiliation{Centre for Quantum Computation and Communication Technology, School of Electrical Engineering and Telecommunications, The University of New South Wales, Sydney, New South Wales 2052, Australia}

\author{Menno Veldhorst}
\affiliation{QuTech and Kavli Institute of Nanoscience, TU Delft, Lorentzweg 1, 2628CJ Delft, the Netherlands}

\author{J.C.C. Hwang}
\affiliation{Centre for Quantum Computation and Communication Technology, School of Electrical Engineering and Telecommunications, The University of New South Wales, Sydney, New South Wales 2052, Australia}

\author{C. H. Yang}
\affiliation{Centre for Quantum Computation and Communication Technology, School of Electrical Engineering and Telecommunications, The University of New South Wales, Sydney, New South Wales 2052, Australia}

\author{Gerhard Klimeck}
\affiliation{Network for Computational Nanotechnology, Purdue University, West Lafayette, IN 47907, USA}

\author{Andrea Morello}
\affiliation{Centre for Quantum Computation and Communication Technology, School of Electrical Engineering and Telecommunications, The University of New South Wales, Sydney, New South Wales 2052, Australia}

\author{Andrew S. Dzurak}
\affiliation{Centre for Quantum Computation and Communication Technology, School of Electrical Engineering and Telecommunications, The University of New South Wales, Sydney, New South Wales 2052, Australia}

\author{Rajib Rahman}
\affiliation{Network for Computational Nanotechnology, Purdue University, West Lafayette, IN 47907, USA}

\date{\today}

\maketitle 

\textbf{S1. Analytic effective mass model to explain the effect of spin-orbit interaction (SOI) on the $g$-factors in a silicon (Si) quantum dot (QD)}

Here we briefly summarize the effective mass model that qualitatively explains the atomistic tight-binding results in terms of the Rashba and Dresselhaus SOI, shown in the supplementary of ref.\ \cite{rifat_arxiv_2016}. We can write the electron Hamiltonian in a Si QD as,

\begin{align}
H=\frac{{{\hbar }^{2}}}{2m}{{\mathbf{k}}^{2}}+V(\mathbf{r})+{{H}_{\textup{Z}}}+{{H}_{\textup{SO}}}
\label{equ1}
\end{align}

\noindent
Here, $m$ is the electron effective mass, $V(\mathbf{r})$ is the potential defining the quantum dot. ${{H}_{Z}}=\frac{1}{2}g\mu \bm{\sigma }\cdot\mathbf{B}$ is the Zeeman term, with $g$ the electron $g$-factor, $\mu_\textup{B}$ the Bohr magneton, and $\mathbf{B}$ the applied magnetic field. ${{H}_{\textup{SO}}}=\beta \left( {{\sigma }_{x}}{{k}_{x}}-{{\sigma }_{y}}{{k}_{y}} \right)+\alpha \left( {{\sigma }_{x}}{{k}_{y}}-{{\sigma }_{y}}{{k}_{x}} \right)$ is the SOI term, where $\beta$ ($\alpha$) is the strength of Dresselhaus (Rashba) interaction, $\sigma_x$, $\sigma_y$ are the Pauli spin matrices, and $k_x$, $k_y$ are electron canonical momentum along x ([100]) and y ([010]) directions respectively. 
$\mathbf{k}=-i\bm{\nabla }-e\frac{{\mathbf{A}}}{\hbar }$, where $\mathbf{A}$ is the vector potential and $\mathbf{B}=\bm{\nabla }\times \mathbf{A}$. Now, we treat ${{H}_{\textup{P}}}={{H}_{\textup{Z}}}+{{H}_{\textup{SO}}}$ as a perturbation to ${{H}_{\textup{0}}}=\frac{{{\hbar }^{2}}}{2m}{{\mathbf{k}}^{2}}+V(\mathbf{r})$. The unperturbed Hamiltonian ${{H}_{0}}$ yields the spin degenerate eigenstates of a Si QD. 


We can write the perturbation Hamiltonian separately for each valley as,

\begin{multline}
{{H}_{\textup{P}}}(v_{\pm })={{\sigma }_{x}}\left( \frac{1}{2}g\mu_\textup{B} {{B}_{x}}+\beta_{\pm } {{k}_{x}}+\alpha_{\pm } {{k}_{y}} \right)+\\
{{\sigma }_{y}}\left( \frac{1}{2}g\mu_\textup{B} {{B}_{y}}-\beta_{\pm } {{k}_{y}}-\alpha_{\pm } {{k}_{x}} \right)+{{\sigma }_{z}}\left( \frac{1}{2}g\mu_\textup{B} {{B}_{z}} \right)
\label{equ2}
\end{multline}
\noindent
Here, we replaced $\beta ,\alpha $ with ${{\beta }_{\pm }},{{\alpha }_{\pm }}$ to denote the two $v_{+}$ and $v_{-}$ valley states \cite{veldhorst_prbrap_2015,nestoklon_prb_2008}. 
Now, the momentum operator in a magnetic field $\mathbf{B}$ are,

\begin{align}
{{k}_{x}}=-i\frac{\partial }{\partial x}-e\frac{{{A}_{x}}}{\hbar }\approx-e\frac{{{A}_{x}}}{\hbar }\\
{{k}_{y}}=-i\frac{\partial }{\partial y}-e\frac{{{A}_{y}}}{\hbar }\approx-e\frac{{{A}_{y}}}{\hbar }
\label{equ4}
\end{align} 
\noindent
Since $\left\langle v_{\pm } \right|-i\frac{\partial }{\partial x}\left| v_{\pm } \right\rangle \approx \left\langle  v_{\pm } \right|-i\frac{\partial }{\partial y}\left| v_{\pm } \right\rangle \approx 0$. For a magnetic field in the x-y plane, we assume, ${{A}_{z}}=0$, ${{A}_{x}}=z{{B}_{y}}$ and ${{A}_{y}}=-z{{B}_{x}}$, where $B_x$ and $B_y$ are the x and y components of the magnetic field respectively and the z component $B_z=0$. Then,

\begin{align}
&{{k}_{x}}\approx \left| e \right|\frac{z}{\hbar }{{B}_{y}}\\
&{{k}_{y}}\approx -\left| e \right|\frac{z}{\hbar }{{B}_{x}}
\label{equ5}
\end{align}

and,
\begin{multline}
{{H}_{\textup{P}}}(v_{\pm})={{\sigma }_{x}}\left( \frac{1}{2}g\mu_\textup{B} {{B}_{x}}+\beta_{\pm } \left| e \right|\frac{z}{\hbar }{{B}_{y}}-\alpha_{\pm } \left| e \right|\frac{z}{\hbar }{{B}_{x}} \right)+\\
{{\sigma }_{y}}\left( \frac{1}{2}g\mu_\textup{B} {{B}_{y}}+\beta_{\pm } \left| e \right|\frac{z}{\hbar }{{B}_{x}}-\alpha_{\pm } \left| e \right|\frac{z}{\hbar }{{B}_{y}} \right)
\label{equ6}
\end{multline}

It is interesting to note here that, when $B_{\textup{ext}}$ is along [100], $B_y=0$ and the Dresselhaus SOI has only off-diagonal contribution ($\sigma_y$), and thus has negligible effect to the $g$-factors. On the other hand, for $B_{\textup{ext}}$ along [110], $B_x=B_y$ and the Dresselhaus SOI acts as a diagonal term and dominates the $g$-factor renormalization. 

After diagonalizing \ref{equ6} we get the spin splitting for different valleys,

\begin{multline}
{{E}_{{\textup{ZS}}(\pm )}}=2\Bigg\{ {{\left( \frac{1}{2}{{g}_{\bot }}\mu_\textup{B} {{B}_{x}}+{{\beta }_{\pm }}\left| e \right|\frac{\left\langle z \right\rangle }{\hbar }{{B}_{y}}-{{\alpha }_{\pm }}\left| e \right|\frac{\left\langle z \right\rangle }{\hbar }{{B}_{x}} \right)}^{2}}+\\
{{\left( \frac{1}{2}{{g}_{\bot }}\mu_\textup{B} {{B}_{y}}+{{\beta }_{\pm }}\left| e \right|\frac{\left\langle z \right\rangle }{\hbar }{{B}_{x}}-{{\alpha }_{\pm }}\left| e \right|\frac{\left\langle z \right\rangle }{\hbar }{{B}_{y}} \right)}^{2}}\Bigg\} ^{\frac{1}{2}}
\label{equ7}
\end{multline}

\noindent
We replaced $g$ here with ${{g}_{\bot }}$, the g-factor perpendicular to the valley axis\cite{roth_pr_1960}. When $B_{\textup{ext}}$ creates an angle $\phi$ with the [100] crystal orientation in a counter clockwise rotation, $B_x=B_{\textup{ext}}cos\phi$ and $B_y=B_{\textup{ext}}sin\phi$,

\begin{multline}
{{E}_{{\textup{ZS}}(\pm )}}={{g}_{\bot }}\mu_\textup{B} B_{\textup{ext}}\Bigg\{1+4{{\left( \frac{\left| e \right|\left\langle z \right\rangle }{{{g}_{\bot }}\mu_\textup{B} \hbar } \right)}^{2}}\left( {{\beta }_{\pm }}^{2}+{{\alpha }_{\pm }}^{2} \right)-\\
4\frac{\left| e \right|\left\langle z \right\rangle }{{{g}_{\bot }}\mu_\textup{B} \hbar }{{\alpha }_{\pm }}+4\frac{\left| e \right|\left\langle z \right\rangle }{{{g}_{\bot }}\mu_\textup{B} \hbar }{{\beta }_{\pm }}\sin 2\phi -\\
8{{\left( \frac{\left| e \right|\left\langle z \right\rangle }{{{g}_{\bot }}\mu_\textup{B} \hbar } \right)}^{2}}{{\beta }_{\pm }}{{\alpha }_{\pm }}\sin 2\phi\Bigg\}^{\frac{1}{2}}
\label{equ8}
\end{multline}

\noindent
Now, $1 \gg \left( \frac{\left| e \right|\left\langle z \right\rangle }{g\mu_\textup{B} \hbar } \right){{\beta }_{\pm }}> \left( \frac{\left| e \right|\left\langle z \right\rangle }{g\mu_\textup{B} \hbar } \right){{\alpha }_{\pm }}$ \cite{nestoklon_prb_2008}. So, we can ignore the second order terms. Thus simplifying equation \ref{equ8} we get,
\begin{align}
{{E}_{{\textup{ZS}}(\pm )}}={{g}_{\bot }}\mu_\textup{B} {B_{\textup{ext}}}{{\left\{ 1-4\frac{\left| e \right|\left\langle z \right\rangle }{{{g}_{\bot }}\mu_\textup{B} \hbar }{{\alpha }_{\pm }}+4\frac{\left| e \right|\left\langle z \right\rangle }{{{g}_{\bot }}\mu_\textup{B} \hbar }{{\beta }_{\pm }}\sin 2\phi  \right\}}^{\frac{1}{2}}}
\label{equ9}
\end{align}

\noindent
After doing a series expansion and ignoring higher order terms, we can simplify this expression even further,

\begin{align}
{{E}_{{\textup{ZS}}(\pm )}}={{g}_{\bot }}\mu_\textup{B} {B_{\textup{ext}}}\left\{ 1-2\frac{\left| e \right|\left\langle z \right\rangle }{{{g}_{\bot }}\mu_\textup{B} \hbar }{{\alpha }_{\pm }}+2\frac{\left| e \right|\left\langle z \right\rangle }{{{g}_{\bot }}\mu_\textup{B} \hbar }{{\beta }_{\pm }}\sin 2\phi  \right\}
\label{equ10}
\end{align}

Writing, $E_{\textup{ZS}(\pm )}=(g_{\bot }+\delta g_{\pm})\mu_\textup{B} B_{\textup{ext}}$, we get,
\begin{align}
\delta g_{\pm}\approx2\frac{\left| e \right|\left\langle z \right\rangle }{\mu_\textup{B} \hbar }\left(-{{\alpha }_{\pm }}+{{\beta }_{\pm }}\sin 2\phi \right) 
\label{equ11}
\end{align} 


\textbf{S2. Ramsey experiments}

One-electron and three-electron Ramsey oscillations for $Q_1$, $Q_2$ and $Q_3$ are shown Fig.\ S1. The dephasing times shown in Fig.\ 2d of the main text, are extracted from Fig.\ S1.

\begin{figure}[htbp]
\includegraphics[width=\linewidth]{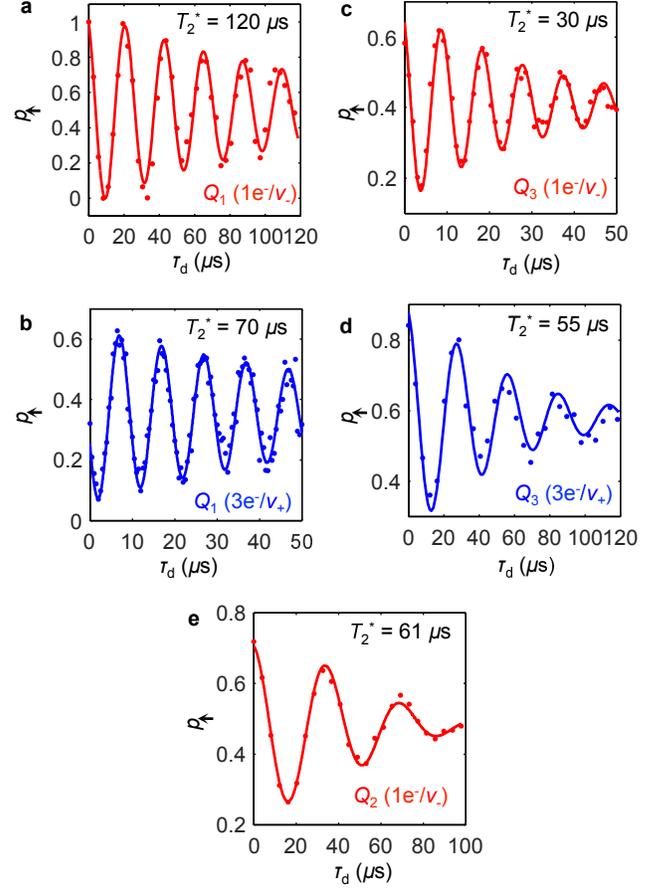}
\caption{Ramsey oscillations for both one-electron ($v_-$) and three-electron ($v_+$) qubits in $Q_1$ and $Q_3$, and one-electron ($v_-$) qubit for $Q_2$ in sample A}
\vspace{0cm}
\label{fi1}
\end{figure}

\clearpage

 \textbf{S3. g-factors in a Si QD with the external magnetic field along the [100] crystal orientation}
 
\begin{figure}[htbp]
\includegraphics[width=\linewidth]{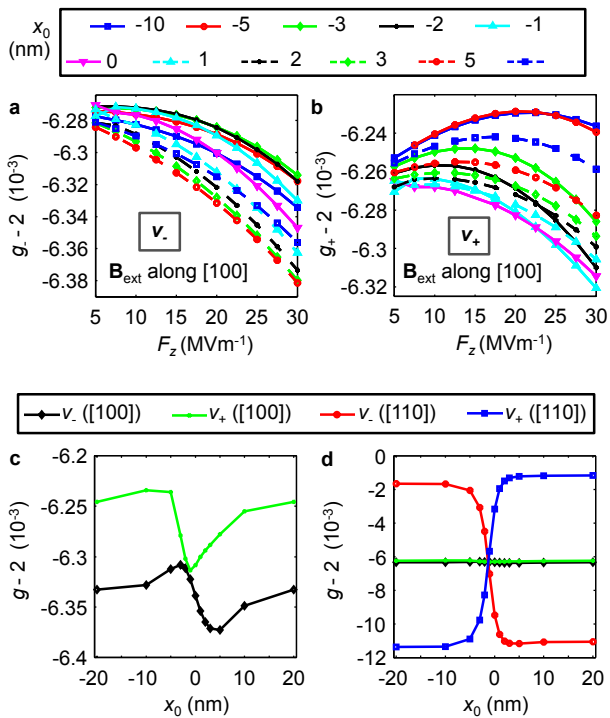}
\caption{Change in $g_-$ (a) and $g_+$ (b) as a function of $F_z$ for different distances from the interface step to the dot center, $x_0$, with $B_{\textup{ext}}$ along [100]. (d) $g_{\pm}$ as a function of $x_0$ for $B_{\textup{ext}}$ along [100], for $F_z$=28.5 MVm$^{-1}$. (c) Comparison in $g_{\pm}$ as a function of $x_0$ between $B_{\textup{ext}}$ along [100] and [110], for $F_z$=28.5 MVm$^{-1}$. }
\vspace{0cm}
\label{fi2}
\end{figure}  

In Fig.\ S2 we show the changes in the $g$-factors with interface step position and electric field for $B_{\textup{ext}}$ along [100]. Figs.\ S2a and S2b show $g_-$ and $g_+$ respectively as a function of $F_z$ for different values of $x_0$. It can be seen that the range of change with both $F_z$ and $x_0$ is much smaller compared to that in Figs.\ 1c and 1d of main text, when $B_{\textup{ext}}$ is along [110]. In Fig.\ S2d we compare the change in $g$-factors between [100] and [110] with respect to $x_0$ for $F_z=$28.5 MVm$^{-1}$. Though there are variations in $g_-$ and $g_+$ due to the change in the location of the dot from an interface step when $B_{\textup{ext}}$ is along [100], as shown by Fig.\ S2c, these variations are negligible compared to that when $B_{\textup{ext}}$ is along [110].


%